# Machine Learning Techniques for Stellar Light Curve Classification

Trisha A. Hinners[1,#], Kevin Tat[1,2], and Rachel Thorp[1,3]

[1]*Northrop Grumman Corporation, One Space Park, Redondo Beach, CA 90278*

[2]*California Institute of Technology, Pasadena, CA*

[3]*University of California Berkeley, Berkeley, CA*

## ABSTRACT

We apply machine learning techniques in an attempt to predict and classify stellar properties from noisy and sparse time series data. We preprocessed over 94 GB of Kepler light curves from MAST to classify according to ten distinct physical properties using both representation learning and feature engineering approaches. Studies using machine learning in the field have been primarily done on simulated data, making our study one of the first to use real light curve data for machine learning approaches. We tuned our data using previous work with simulated data as a template and achieved mixed results between the two approaches. Representation learning using a Long Short-Term Memory (LSTM) Recurrent Neural Network (RNN) produced no successful predictions, but our work with feature engineering was successful for both classification and regression. In particular, we were able to achieve values for stellar density, stellar radius, and effective temperature with low error (~ 2 - 4%) and good accuracy (~ 75%) for classifying the number of transits for a given star. The results show promise for improvement for both approaches upon using larger datasets with a larger minority class. This work has the potential to provide a foundation for future tools and techniques to aid in the analysis of astrophysical data.

## 1 INTRODUCTION

Future space-based telescopes and ground-based observatories have a potential to add a large amount of unprocessed data into the astronomy community in the coming decade. For example, the Hubble Space Telescope (HST) produced approximately 3GB per day, whereas the James Webb Space Telescope (JWST) is expected to produce approximately 57.5 GB per day (Beichman, 2014). Taking this to further extremes, the Square Kilometer Array (SKA) which will be online in 2020 is predicted to produce on the order of $10^9$ GB per day; this is the same amount of data the entire planet generates in a year (Spencer, 2013). Recent advances in computer science, particularly data science, have the potential to not only allow the astronomy community to make predictions about their data quickly and accurately but also to potentially aid in discovering which features make objects distinguishable. These features may or may not be known by the human analyst and the method could have the potential to discover a relationship within the data previously unknown by the human astronomer.

---

[#] All correspondence to Trisha A. Hinners, Northrop Grumman Corporation, Redondo Beach, CA 90278, USA, trisha.hinners@ngc.com

There have been a number of efforts to extract meaning from stellar light curves over the past few years. Most extract specific features from a curve to tackle one particular physical property or come up with novel data processing techniques to improve analysis through a reduction of noise and/or variability. Some notable examples include work by Richards et al. 2011 using periodicity features to measure stellar variability, Bastien et al. 2015 to extract flicker to measure stellar surface gravity, and Wang et al. 2016 used data driven-models with pixel-level de-trending to produce low-noise light curves while retaining important transit signals.

With the recent interest in machine learning, we decided to approach the problem of understanding what physical properties of a star are most related to the light curve using two complementary machine learning approaches: feature engineering and representation learning. Feature engineering takes raw data and summarizes that data with features that are deemed important by the analyst. These features are then fed into a machine learning method. Representation learning differs from feature engineering in that the machine learning method is allowed to learn what attributes best distinguish the data, removing the bias from the analyst.

There are very few examples using machine learning techniques in astronomy, but that number is growing. One of the first examples dates back to 2007 with Bailey et al. doing object classification for supernovae using the Supernovae Factory data with synthetic supernovae as training data. In 2010 Ball et al, published a review paper on the uses of machine learning methods in astronomy. More recent examples include work by Armstrong et al. (2017) on transit shapes and Thompson et al. 2015 on transit metrics, both using real Kepler data. Thompson et al. described a new metric that uses machine learning to determine if a periodic signal found in a photometric time series appears to be transit shaped. Using this technique they were able to remove 90% of the non-transiting signals and retain over 99% of the known planet candidates. This study was done with feature engineering and extraction methods.

Examples from the supernovae community include work on both real and simulated data. Cabrera-Vives et al. (2017) used a convolutional neural network (CNN) for classifying images of transient candidates into either artifacts or real sources. Their training data set included both real transients and simulated transients. They were able to distinguish between real and fake transients with high accuracy. Both Karpenka et al. 2013 and Charnock & Moss 2017 used deep learning approaches on simulated supernovae light curves from the SuperNova Photometric Classification Challenge (SNPCC). Karpenka et al. used a perceptron artificial neural network for binary supernovae classification. The perceptron is a supervised learning method based on a linear predictor function. Charnock & Moss used an LSTM RNN to classify the synthetic supernova with a high rate of success. Their dataset consisted of just over 21,000 synthetic supernovae light curves. This work inspired us to use an LSTM RNN approach as our first attempt at applying machine learning to stellar light curve classification for the purpose of characterizing host stars.

The work described in this paper is divided into two separate efforts; an approach in representation learning and an approach in feature engineering. For our representation learning efforts we utilize a bi-directional Long Short-Term Memory (LSTM) Recurrent Neural Network (RNN) to both predict and classify properties from Kepler light curves. For the feature engineering approach we utilize a Python library called FATS (Feature Engineering for Time Series) (Nun et al 2015) which facilitates and standardizes feature extraction for time series data, and was specifically built for astronomical light curve analysis. To the best of our knowledge this is the first work to do a comparative study of representation learning and feature engineering for prediction and classification using real astronomical data of light curves.

## 2 DATA

In an attempt to make this study widely applicable, we classify a large number of Kepler object light curves according to a wide range of stellar properties. The respective sources and formats of both the time series measurements and property labels are discussed below.

### 2.1 Light Curves

All light curves used in this study were from the Mikulski Archive for Space Telescopes (MAST)[1]. The physical parameters[2] and their descriptions[3] were obtained from the table of stellar properties using the *Kepler_stellar17.csv.gz* file. Kepler flux measurements were made over multiple quarters for each source, where the instrument rotates by 90 degrees from one quarter to the next to re-orient its solar panels. The quarters are approximately 90 days long, with a data sampling every 29.4 minutes for long cadence observations and every 58.8 seconds for short cadence observations. Out of the 200,000 total stars, only 512 are short cadence. In order to maintain consistent data sample structures, the short-cadence light curves are removed from the dataset. This is consistent with common practice.

We iteratively ran through the archived Kepler quarter files and downloaded more than 234,000 files (~94 GB). The files are formatted as Flexible Image Transport System (FITS) files (most commonly used digital file format in astronomy), which contain headers that describe the observing environment and data quality and a table of the flux measurements over time. There are two values reported for the flux measurements: Simple Aperture Photometry (SAP) flux and Pre-Search Data Conditioning (PDC) SAP flux. The PDC SAP versions of the light curves remove instrumental variations and noise in the data, while preserving both stellar and transiting exoplanet behavior. Therefore this is the flux measurement used to construct the light curves.

Recall that the header of the light curve file contains information about the quality of data. In an attempt to keep only observations with reliable signals, we filter out all quarters that contain either:

- Contamination from neighboring stars greater than 5% of the total measured flux, or
- A flux yield less than 90% of that object's total flux

Before training the remaining data, a number of preprocessing steps are required. These are given in order below:

1. Keep every tenth data point to make the files sparser, cutting down on computation time. The initial Kepler data is extremely dense. We found by visual inspection that sampling every 10th data point still yielded representative curves while allowing for a larger number of targets. Using all time steps for each target was too slow to be tractable; thinning it out

---

[1] https://mast.stsci.edu
[2] https://archive.stsci.edu/kepler/catalogs.html
[3] http://archive.stsci.edu/search_fields.php?mission=kepler_stellar17

allowed us to include enough diversified targets to be an effective attempt at machine learning

2. Normalize the curve. Raw flux values contain relatively little information on their own and are extremely inconsistent across a single object's multiple quarters.
   a. Divide each by the median of the curve.
   b. Subtract one from all points to shift down and center the curve about zero.
3. Iterative $2.5\sigma$ clipping ($\sigma$ = 1 standard deviation) to remove extreme outliers from the data (likely to be remaining instrumental artifacts).
4. Pseudo-random data augmentation to fill gaps in the data, preserving the time step information. In particular, we identify any consecutive gap that exists in the data (denoted in the files with NaNs = "Not a Number") and fill each missing time slice with a random value between the two real values on either side of that respective gap. To do this we first group together sequential NaN appearances (which is one or more NaN entries bounded by real flux values on either side). Next for each of these empty values we substitute a random value between the left and right flux values on either side of the corresponding gap. Since the NaN values are not measurements that we can assume to know (i.e. missing data) we wanted to avoid imposing any interpolated trend/behavior that may or may not be present. Ideally, the model will learn to ignore these noisy, random portions, as if the data was not present.

In Figure 1, we display an example light curve quarter before and after preprocessing has been performed, respectively.

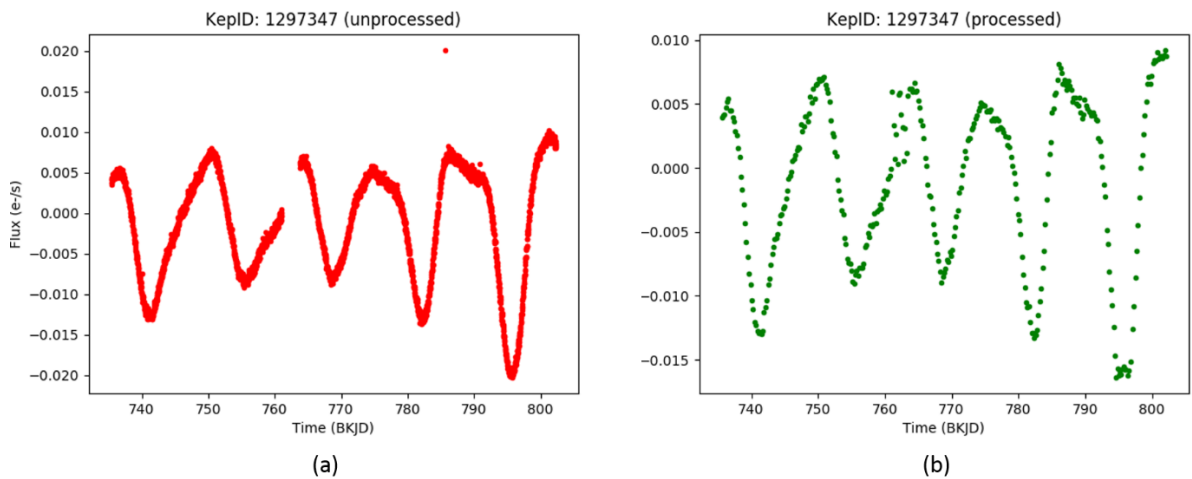

Figure 1 – A demonstration of before (a) and after (b) light curve preprocessing on a single quarter

Finally, we concatenate all processed quarters of a single object into one large curve. To properly format the data for both the RNN and feature engineering approaches, all curves must be the same length; therefore we find the longest resulting light curve and prepend all others with -999s until they are the same length. The -999 value is distinct and is masked out later in the training and testing process.

This process reduced the initial 234,000 quarter files down to just over 48,500 unique object light curves, each with a length of ~7,000 time slices.

## 2.2 Labels

The stellar properties used as labels in the classification and regression tasks were extracted from the Kepler Stellar 17 table on MAST[4]. This table includes properties for more than 200,000 Kepler targets, and of the 95 columns describing each target[5], we use the 10 properties given in Table 1 to generate labels for 10 distinct prediction tasks.

| Parameter Name | Description | Units | Minimum | Maximum | Prediction Task |
|---|---|---|---|---|---|
| teff | Effective Temperature | K | 2500 | 27730 | Regression |
| logg | Surface Gravity | $\log_{10}(cm/s^2)$ | 0.016 | 5.52 | Regression |
| feh | Metallicity | dex | -2.5 | 1 | Regression |
| mass | Mass | $M_\odot$ | 0.09 | 3.74 | Regression |
| radius | Radius | $R_\odot$ | 0.104 | 300.749 | Regression |
| dens | Density | $g/cm^3$ | 0 | 124 | Regression |
| kepmag | Kepler-band Magnitude | mag | -0.419 | 17.394 | Regression |
| nconfp | Number of confirmed planets | - | 0 | 7 | Classification |
| nkoi | Number of associated KOIs[6] | - | 0 | 7 | Classification |
| ntce | Number of associated TCEs[7] | - | 0 | 8 | Classification |

Table 1 - Ten stellar properties used to generate labels for corresponding object light curves in prediction tasks.

## 3 METHOD

As stated in the introduction, we approach the problem of extracting and identifying physical properties of the star through two methods: representation learning and feature engineering. In the sections below, we describe how each method was implemented. Results and discussion for each approach will follow in subsequent sections.

### 3.1 Machine Learning Introduction

Machine learning attempts to automate the data analysis process. Much of this is done by exploiting the tools of probability theory. There are many different flavors of machine learning, but it is usually divided into two main types. In predictive or supervised learning, the goal is to learn a mapping from inputs to outputs given a labeled set of input-output pairs (the training set). The second main type is descriptive or unsupervised learning where we are only given inputs and the goal is to find interesting patterns in the data. Within this space one can perform either classification (pattern recognition) when the problem is categorical or regression to find a specific value. (Murphy 2012). There are a large variety of algorithm approaches to machine learning.

---
[4] Found at: https://archive.stsci.edu/pub/kepler/catalogs/
[5] Parameter information further discussed at: http://archive.stsci.edu/kepler/stellar17/help/columns.html
[6] Kepler Objects of Interest
[7] Threshold Crossing Events, e.g. exoplanet transits

These include (as a few examples) Bayesian, clustering, ensemble, instance based, artificial neural networks, regularization, and feature engineering.

Artificial neural networks (ANNs) are a much-lauded tool of machine learning, popular for their flexibility and power. ANNs can be thought of mathematically as a form of function approximation and are used for tasks such as regression and classification. ANN's have a large number of tunable parameters, all of which fall into two categories. The weights and biases are internal parameters, selected via an optimization routine (e.g. stochastic gradient descent) over a chosen metric (e.g. mean squared error), Bottou 2010. The hyperparameters include width, which is the number of nodes per layer, and depth, which is the number of layers stacked to form the network. Increasing the depth leads to *deep learning*, where the definition of *deep learning* tends to change with advances in computation. In general, increasing depth and width enables better performance, where depth is thought to have a greater payoff. Note that as depth and width increase, so does computation time.

Every ANN consists of nodes and edges, with an example ANN shown in Figure 2. The nodes of the input layer correspond to the input data; the number of input nodes corresponds to the dimensionality of that data. For example if we are interested in 8x8 pixel grayscale images, the number of input nodes will be 64. The nodes of the output layer depend on the task for which the ANN was designed. If the network was designed to predict a single number, then there were be a single output node. If the intent is to classify an input image among *k* categories, one would choose *k* output nodes, each corresponding to the probability the input data belongs to a particular category. The nodes of the hidden layer correspond to intermediate data transformations, which are governed by both the learned weights and biases, and the hyperparameters. The arrows in Figure 2 denote the flow of data between nodes; the example is a densely connected feed-forward network, where data is allowed to flow between all nodes (densely connected), but only in the one direction (feed-forward).

The details of each individual node are shown in Figure 2 on the diagram on the right. Each node takes some number of inputs, combines them, feeds them through an activation function, and the result is then passed on to some other number of nodes. The combination of inputs is usually scaled by individual weights, then added along with a bias term. The weights and biases are chosen via optimization but the activation function is chosen a priori. The choice of activation function is a topic of active research but a current popular choice is the rectifier (Glorot et al. 2011).

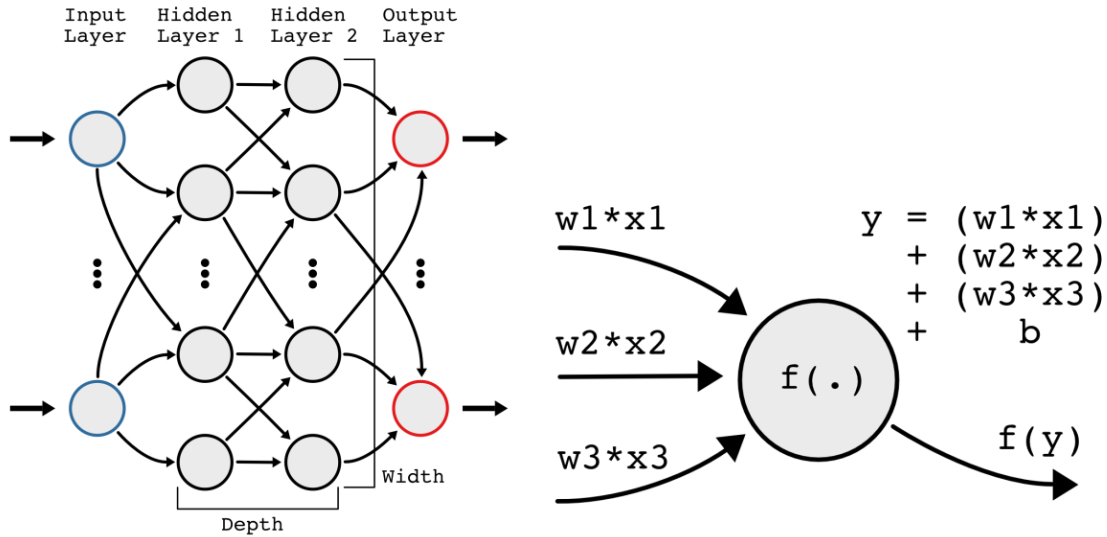

Figure 2 – The figure on the left is a schematic for a simple artificial neural network with a depth of two and unspecified width and the figure on the right is a schematic for a simple, generic node. Here three inputs $x_i$ are scaled by weights $w_i$, summed, and biased b before being fed through an activation function *f(y)*.

There are several flavors of ANN's but two of the most popular are Recurrent Neural Networks (RNNs) and Convolutional Neural Networks (CNNs). An RNN is an ANN with internal memory. What this means is that information is allowed to pass between nodes in the same layer. If both forward and backward propagation are allowed, then the RNN becomes bi-directional. A popular form of RNNs is the Long Short-Term Memory (LSTM) RNN which provides facilities for memory management including the ability to forget or reset an internal state (Sak et al. 2014). A CNN is an ANN which assumes some invariance structure of the input data, and therefore enforces invariance in the structure of the ANN (Zhang et al. 1990). The network is invariant in the sense that it applies the same small set of weights and biases to different portions of the input data. More specifically, the network performs convolution of learned kernels against the input data. This design choice reduces the number of internal parameters, decreasing the expense of training. Convolving against a kernel can be thought of as searching for patterns. Mishkin et al. 2016 details an investigation of different CNN design choices and their relative performance.

A more thorough discussion about machine learning in general can be found in Murphy 2012 and neural networks in particular in the review article by Schmidhuber 2015.

### 3.2 Representation Learning

In an effort to understand which properties we can obtain from stellar light curves, we turn to representation learning techniques. Representation learning allows the model to extract the "features" that it finds to be important in characterizing objects according to one physical property at a time. While this may initially limit model interpretability, it provides an opportunity for hidden features of the light curve to surface and help in classifying an object by various stellar properties. While feature engineering can be extensive, representation learning has the opportunity to remove human-based preconceived notions about what may or may not affect a star's classification.

In line with more common natural language processing (NLP) tasks, we treat each set of 48,500 light curves as a corpus, with each light curve simulating a sentence and each normalized flux measurement simulating a word. By implementing a LSTM RNN we hope that the model will learn both semantic relations between flux values in a sequence, as well as the more general pattern meanings throughout the "corpus" of objects, allowing the model to make accurate stellar predictions. A similar approach of treating a light curve as a sentence was done by Charnock & Moss 2017 in their analysis of supernovae light curves.

### 3.2.1 Network Architecture

We referred to related literature when determining the model architecture – primarily Charnock & Moss 2017, which optimized a LSTM RNN in classifying supernovae light curves. Aside from similar data structures, Charnock & Moss had a comparable data set size (although simulated) and analogous prediction tasks, leading us to utilize a similar model structure and complexity. Our LSTM RNN was built in python using the Keras neural network library[8]. The ideal architecture was found to be an RNN with two LSTM hidden layers of 16 nodes each (and an initial masking layer to filter out prepended -999s), although we initially ran tests on just a single hidden layer to reduce computation time. An example of a generic RNN with two LSTM hidden layers can be seen in Figure 3.

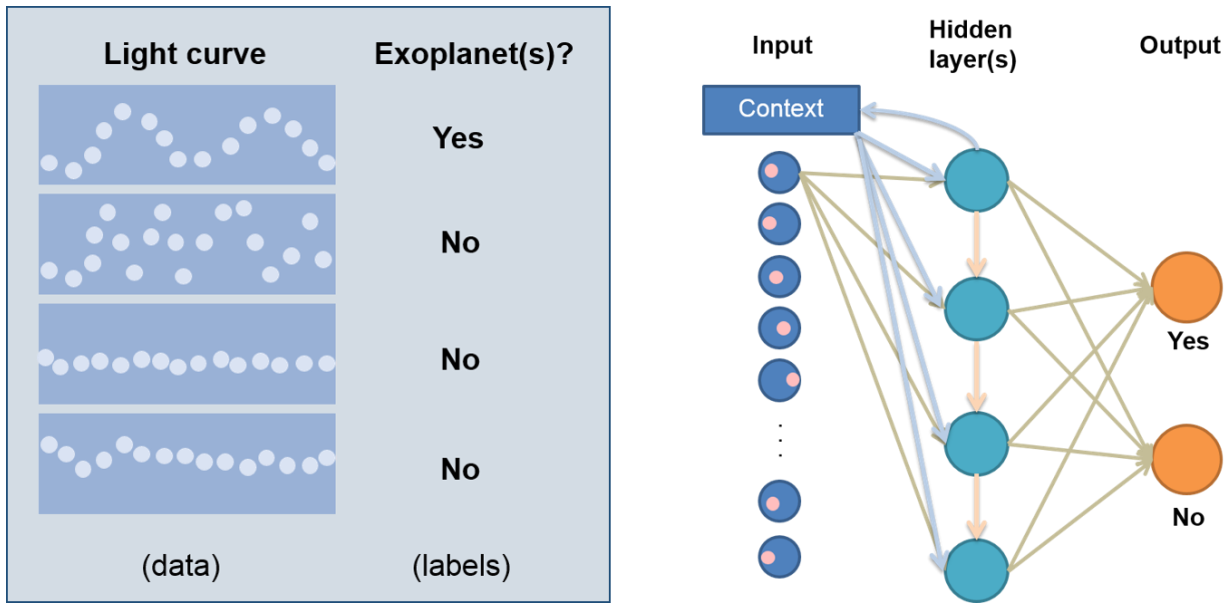

Figure 3 - An example of the data and label inputs and architecture of a recurrent neural network with LSTM nodes predict and classify stellar properties

In building the LSTM RNN, we wanted to perform two types of predictive tasks: binary classification (*does an object belong to one class or another?*) and regression (*what is the numerical value of this star's physical property?*). For both LSTM layers we used a softmax activation function for classification tasks and a softsign activation function for regression tasks. The dense layer was always assigned a linear activation function. In network compilation, a categorical cross-entropy loss was used for classification tasks and a mean squared error loss for regression tasks. We applied the RMSProp[9] optimizer with default learning rate. When fitting, a batch size of 20 was used and in both classification and regression tasks and we surveyed the reported loss values between epochs to determine when performance had plateaued (namely, when loss values seemed to converge). Weights were balanced using scikit-learn's *class_weight* utility function specifying 'balanced' weights. This process effectively returns the frequency of each class, or more specifically each class weight = (number of samples)/(number of classes * number of samples belonging to that class). Once a list of class weights were obtained, we fed the list into the neural network when fitting. All other parameters not specified here were Keras layer defaults. More specifics on these modifications for the two tasks are discussed below.

---

[8]The Keras deep learning library is an open source library developed by Francois Chollet at MIT in 2015. More information can be found at: https://keras.io/

### 3.2.2 Modifications

We decided to perform classification tasks on the parameters with discrete values within a limited set (i.e. number of confirmed planets, number of associated Kepler Objects of Interest (KOIs), and number of associated Threshold Crossing Events (TCEs)). However, while the sets of possible values for each of these properties are already quite limited, each of these three parameters is heavily dominated by negative signals (i.e. a value of zero). Thus, we decided to simplify the tasks further by making each binary classification task either equal to or greater than zero. However, despite the classification simplification, the data still consisted of a skewed population for each of the three relevant properties. To combat heavy bias, we instantiated the model with balanced weights, such that the model would weight the importance of positive labels more highly than negative labels to avoid converging as a simple majority class predictor.

The key to creating a classification model versus a model that performs regression tasks is primarily in the structure of the output layer. We implemented one node for each class of the classification task (two for binary classification) in the output layer. Additionally, the loss function (a metric over which each model is optimized) varies slightly between classification and regression tasks. Since we want to correctly categorize the target, we apply a categorical cross entropy loss function, provided below:

$$L_i = -\sum_j t_{i,j}\log(p_{i,j})$$

The loss function is applied to each prediction target *i* and each of the *j* possible classes (e.g., *j = 2* in our binary predictions), where *t* is the target or actual probability that an object belongs to that class, and *p* is the corresponding predicted probability. This function is minimized by the neural network to optimize prediction tasks.

Rather than creating multiple nodes in the output layer as was done for the classification tasks, regression tasks require just a single node. Which means that the output will be the estimated property value, rather than a set of weights corresponding to the model's confidence in each respective class.

### 3.2.3 Evaluation Metrics

---

[9] https://keras.io/optimizers/

In determining the success of a predictive model, various evaluation metrics are used to measure results. These are not used in the training process, but are helpful when considering how well the model performed on a certain task. The metrics come from entries in a confusion matrix (Kohavi and Provost, 1998) which contains information about the actual and predicted classifications done by a classification system. Performance of the system, in our case our machine learning methods, is evaluated using the data contained in the confusion matrix.

For a binary classifier we utilize a two class matrix, also known as a truth table as seen in Table 2.

|  | | Actual Value | |
|---|---|---|---|
|  | | 0 | 1 |
| Predicted Value | 0 | TN | FN |
|  | 1 | FP | TP |

*Table 2- Truth Table where TN is the number of True Negatives, FN is the number False Negatives, FP is the number of False Positives, and TP is the number of True Positives. These values are used in calculating the metrics used to evaluate model performance.*

Traditional or "raw" accuracy is simply defined as the ratio between the number of correct predictions and the total number of predictions. For two classes, raw accuracy is calculated as,

$$Accuracy = (TP + TN)/(P + N)$$

where TP is the number of true positives, TN is the number of True Negatives, P is the total number of positives, and N is the total number of negatives.

Therefore, a random binary classifier would have an accuracy of 1/2 and a random classifier with three classes would have an accuracy of 1/3. Each of our classification tasks contained extremely imbalanced data, where the number of "positive" samples (i.e. a confirmed transit) was less than 10% of the total data set for each parameter. Therefore raw accuracy returned misleadingly high values even for a simple majority class predictor and was not an appropriate metric to evaluate prediction performance.

Therefore we turned to balanced accuracy utilizing a confusion matrix of predictions. For binary classification problems, the confusion matrix splits predictions into true positives (TP), false positives (FP), false negatives (FNs), and true negatives (TNs). Balanced accuracy is then defined as;

$$Balanced\ Accuracy = (^{TP}/_P + ^{TN}/_N)/2$$

Additionally we can calculate the recall of a model which tells us how many positive cases were correctly identified and the precision which tells us how many of those predicted positive cases were correctly identified. Recall is defined as:

$$Recall = TP/(TP + FN)$$

and precision is defined as:

$$Precision = TP/(TP + FP)$$

From recall and precision we can calculate the $F_1$ score which can be used to measure the accuracy of the model. It uses both recall and precision (which are obtained from the Truth Table). The $F_1$ score provides a harmonic average of the precision and recall, where the best value for $F_1$ is 1 and the worst is 0. F1 is described as,

$$F_1 = 2 * \frac{Precision * Recall}{Precision + Recall}$$

### 3.2.4 Representation Learning (LSTM RNN) Results

We found that both the classification and regression results resulted in approximately guessing accuracy. For classification, the balanced accuracy was approximately 52% for all three tasks and for regression the RNN was only finding the average of all of the values instead of predicting the individual value. Previous successful work had centered on simulated data (Charnock & Moss) or other types of representation learning (Self Organizing Map (SOM) in the case of Armstrong et al. 2017) and it is possible that the variation in the light curve data, the relatively small sample set of 48,500 light curves, or the ratio of positive samples to negative was too low for the RNN to predict values.

Upon these results we were driven to explore another machine learning method with more human knowledge in the loop – namely feature engineering.

### 3.3 Feature Engineering

To attempt to improve our ability to perform prediction tasks, we turned to feature extraction to construct feature representations of the light curves. This strategy is commonly referred to as feature engineering. Feature engineering is the process of determining, calculating, and extracting features from raw data. These features are typically properties of the raw data that are human interpretable and believed to provide insight on the prediction task at hand. While this process can be arduous (feature engineering often takes quite a lot of time and may be computationally intensive), it also provides useful intuition into our understanding of certain properties and patterns of the raw data. Furthermore, the fact that feature engineering typically utilizes human-crafted properties, allows us to perform a more in-depth data analysis on why certain features predict certain physical properties. While representation learning is able to extract "features" from raw data for unexpected insight, it does not offer the depth of insight that feature engineering can provide.

In order to extract features from light curves, we turned to FATS, Feature Analysis of Time Series (Nun et.al, 2015). Since there is extensive documentation for FATS on its Github repository[10] and within Nun et al. 2015, here we will only summarize how we extracted features from our preprocessed light curves. FATS takes time series data and, depending on the target, extracts mathematical properties and statistical information (Nun et al. 2015). While FATS can be applied to a variety of time series data, we focus on using it to extract features from light curves. Specifically, we extract 46 features from 6038 light curves and train them on the following models: Naïve Bayesian, K-Nearest Neighbors, Support Vector Machine, Decision Tree, Random Forest, L1- Norm Regression, L2-Norm Regression, and Support Vector Regression. This analysis was performed on the same data prepared in the LSTM RNN experiment. Hyperparameters were determined by a simple script that trained from 1 to N classifiers and chose the value that minimized out of sample error. The main data set was split into a testing and training set such that the training set was 80% of the data and the testing set was 20% of the data. Each model used for classification utilized a function to calculate out of sample error and each model used for regression utilized a function that provided RMSE as an output.

### 3.3.1 Feature Selection and Extraction

To accurately simulate the type of information we will be receiving from future synoptic surveys, we utilize only magnitude and time measurements from the light curve as inputs to FATS. Given magnitude and time, there are a total of 53 features that can be calculated. Out of this 53 we exclude seven features. FluxPercentileRatioMid20, FluxPercentileRatioMid35, FluxPercentileRatioMid50, FluxPercentileRatioMid65, and FluxPercentileRatioMid80 were excluded because they produce values of infinity during feature generation. This issue is a product of the preprocessing done on the light curves, which centered each time series around 0. FluxPercentileRatio* were calculated using the formula,

$$FluxPercentileRatio* = \frac{F_{50\pm*/2}}{F_{5,95}}$$

where $F_{5,95}$ is the difference between 95% and 5% of the flux. This difference occasionally truncated to zero by Python if it was too small. This led to values of infinity that were removed from the list of features as they do not provide reliable information. We also discarded Percent Amplitude and Percent Difference Flux Percentile. Percent amplitude is calculated as,

$$Percent\ Amplitude = \max(\frac{F_{min}}{F_{median}}, \frac{F_{max}}{F_{median}})$$

and Percent Difference Flux Percentile from:

$$Percent\ Difference\ Flux\ Percentile = \frac{F_{5,95}}{F_{median}}$$

Both of these values rely on the median flux value, which on occasion is equal to zero as the data was normalized to be centered around zero. Again, during feature generation, some of the light curves held values of infinity for these two properties and so we discarded them with the same reasoning as for the FluxPercentileRatio*values. Therefore we were left with 46 features which are listed in Table 3.

---

[10] https://github.com/isadoranun/FATS

| Feature | Input Data (besides magnitude) | Parameters | Default | Reference |
|---|---|---|---|---|
| Amplitude | - | - | - | Richards et al. (2011) |
| AndersonDarling test | - | - | - | Kim et al. (2008) |
| Autocor length | - | Number of lags | 100 | Kim et al. (2011) |
| Con | - | Consecutive Stars | 3 | Kim et al. (2011) |
| $Eta_e$ | Time | - | - | Kim et al. (2014) |
| Freq1HarmonicsAmp$_0$ | Time | - | - | Richards et. al (2011) |
| Freq1HarmonicsAmp$_i$ | Time | - | - | Richards et. al (2011) |
| Freq1HarmonicsRelPhase$_0$ | Time | - | - | Richards et. al (2011) |
| Freq1HarmonicsRelPhase$_i$ | Time | - | - | Richards et. al (2011) |
| Freq2HarmonicsAmp$_0$ | Time | - | - | Richards et. al (2011) |
| Freq2HarmonicsAmp$_i$ | Time | - | - | Richards et. al (2011) |
| Freq2HarmonicsRelPhase$_0$ | Time | - | - | Richards et. al (2011) |
| Freq2HarmonicsRelPhase$_i$ | Time | - | - | Richards et. al (2011) |
| Freq3HarmonicsAmp$_0$ | Time | - | - | Richards et. al (2011) |
| Freq3HarmonicsAmp$_i$ | Time | - | - | Richards et. al (2011) |
| Freq3HarmonicsRelPhase$_0$ | Time | - | - | Richards et. al (2011) |
| Freq3HarmonicsRelPhase$_i$ | Time | - | - | Richards et. al (2011) |
| Linear Trend | Time | - | - | Richards et. al (2011) |
| Max Slope | Time | - | - | Richards et. al (2011) |
| Mean | - | - | - | Kim et. al (2014) |
| Mean Variance | - | - | - | Kim et. al (2011) |
| Mean Absolute Deviation | - | - | - | Richards et. al (2011) |
| Median BRP | - | - | - | Richards et. al (2011) |
| PairSlopeTrend | - | - | - | Richards et. al (2011) |
| Period Lomb-Scargle | Time | Oversampling Factor | 6 | Kim et al. (2011) |
| Period Fit | Time | - | - | Kim et al. (2011) |
| $\psi_{cs}$ | Time | - | - | Kim et al. (2014) |
| $\psi_\eta$ | Time | - | - | Kim et al. (2014) |
| $Q_{3-1}$ | - | - | - | Kim et al. (2014) |
| RCS | - | - | - | Kim et al. (2011) |
| Skew | - | - | - | Richards et al. (2011) |
| Slotted AutoCor Length | Time | Slot Size T (days) | 4 | Protopapas et al. (2015) |
| Small Kurtosis | - | - | - | Richards et al. (2011) |
| Standard Deviation | - | - | - | Richards et al. (2011) |

Table 3 – Features generated from FATS used in machine learning methods. In the "freqN Harmonics" terms, i = 1, 2, 3.

### 3.3.2 Feature Engineering Regression Results

We ran three different regression models to predict values for stellar surface gravity (log(g)), stellar mass (in units of $M_\odot$), density, stellar radius (in units of $R_\odot$), and the stellar effective temperature.

The first two models are linear regression models. The benefits of linear regression is its simplicity in both implementation and interpretability. Its drawback comes when the relationship between the inputs and outputs cannot be approximated by a linear relationship, in which case the model will give poor predictions. The two linear methods we used are LASSO (least absolute shrinkage and selection operator) (Tibshirani, 1996) and ridge regression (Rasmussen, 2006), which we have denoted as L1 Regression and L2 Regression, respectively. L1 regression is 'robust' meaning it does not overly fit to outliers, it is 'unstable' such that small adjustments in the data have the potential to move the regression fit, and it has multiple solutions. L2 regression is 'not robust' so it could have a tendency to over-fit, it is stable so the regression line is not affected by small data adjustments, and it has a unique solution.

The third method is a non-linear method called Support Vector Regression (SVR). It was originally developed for classification problems and later extended to regression. It is useful when the relationship between the inputs and outputs are not best fit by a linear relationship. For a full description of the methods please refer to Rasmussen et al. 2006 and Murphy 2012. The results obtained for regression are described in terms of Root Mean Squared Error (RMSE) and are shown in Table 4. Recall that RMSE is calculated within each of the regression models.

| Root Mean Squared Error | | | | | |
|---|---|---|---|---|---|
| Model | Stellar Surface Gravity (log(g)) | Stellar Mass ($M_\odot$) | Density (g/cm$^3$) | Stellar Radius ($R_\odot$) | Effective Temperature (K) |
| L1 | ± 0.9088 | ± 0.6604 | ± 2.699 | ± 14.26 | ± 879.6 |
| L2 | ± 0.8254 | ± 0.6341 | ± 2.669 | ± 13.25 | ± 875.4 |
| SVR | ± 0.8735 | ± 0.6050 | ± 2.829 | ± 19.87 | ± 967.2 |

Table 4 - Feature engineering results for regression

Where the range in values for each stellar property is in Table 5.

| | Stellar Surface Gravity (log(g)) | Stellar Mass ($M_\odot$) | Density (g/cm$^3$) | Stellar Radius ($R_\odot$) | Effective Temperature (K) |
|---|---|---|---|---|---|
| Range | 0.016 - 5.52 | 0.09 - 3.74 | 0 - 124 | 0.104 - 300.749 | 2500 - 27730 |

Table 5 – Range of values for each stellar property

To get a better comparison for how each model does at predicting these values, we then normalized the RMS error, which can be seen in Table 6.

| Normalized Root Mean Squared Error (%) | | | | | |
|---|---|---|---|---|---|
| Model | Stellar Surface Gravity (log(g)) | Stellar Mass ($M_\odot$) | Density (g/cm$^3$) | Stellar Radius ($R_\odot$) | Effective Temperature (K) |
| L1 | ± 0.1651 | ± 0.1809 | ± 0.0217 | ± 0.0474 | ± 0.0348 |
| L2 | ± 0.1499 | ± 0.1737 | ± 0.0215 | ± 0.0441 | ± 0.0347 |
| SVR | ± 0.1587 | ± 0.1657 | ± 0.0228 | ± 0.0661 | ± 0.0383 |

Table 6 – Normalized RMS for each of the stellar properties with each different model.

Feature engineering for regression proved to give good predictions for stellar surface gravity and stellar mass, and very good predictions for stellar density, stellar radius, and effective stellar temperature with both of the linear models performing slightly better than SVR. SVR pulls ahead only for predicting the stellar mass and is pretty even with the linear models for predicting stellar surface gravity but overall the differences between the models are minor. This technique could be used with confidence to classify the large database of unclassified stars without immediate need for follow up observations.

### 3.3.3 Feature Engineering Classification Results

Classification was performed on three separate types of events: number of threshold crossing events, number confirmed planets, and number of Kepler objects of interest.

### 3.3.3.1 Feature Importance

Using FATS for each of these classification events we calculated the importance of each of the features from Table 3. Our first look was the number of Kepler objects of interest in a given light curve. As we see in Figure 4, mean, skew, and freq1HarmonicsRel.Phase1 are all features that contribute greatly to understanding if an object of interest is contained within the light curve for a model using a Random Forest classifier.

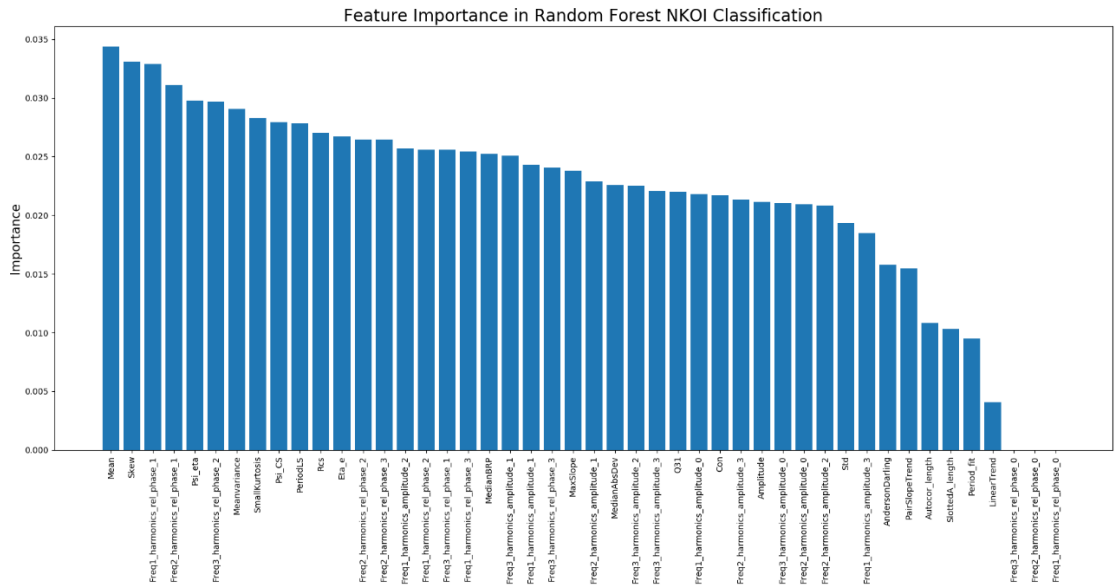

Figure 4 – Feature importance in random forest NKOI classification

Figure 5 shows the relative feature importance for the number of threshold crossing events. Here different features, namely the Period Lomb-Scargle and $Eta_e$ are the two most dominant features but the freqN HarmonicsRel.Phase terms are not important at all to the classification.

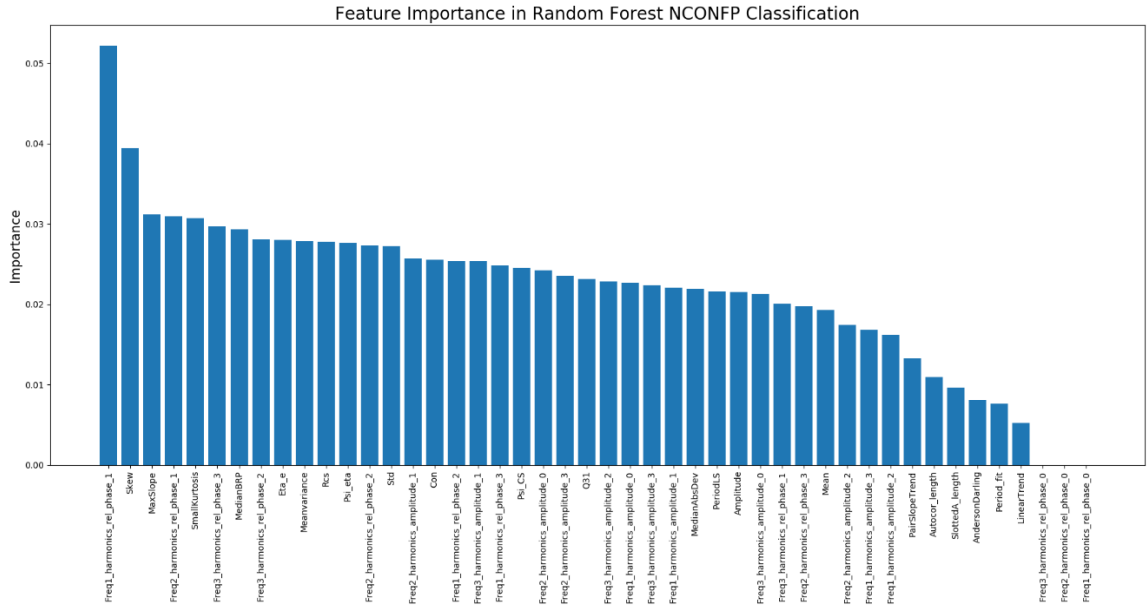

Figure 5 - Feature importance in random forest number of confirmed planets classification

Finally Figure 6, shows the relative feature importance for the number of confirmed planets. Here we see that the freq1_harmonics_ rel_phase_1 is the dominant feature followed by skew, with the features not contributing to the classification being the different freq_harmonics_rel_phase0 terms. Doing this analysis shows us that even for seemingly fairly related events the feature importance can vary greatly when it comes to classification.

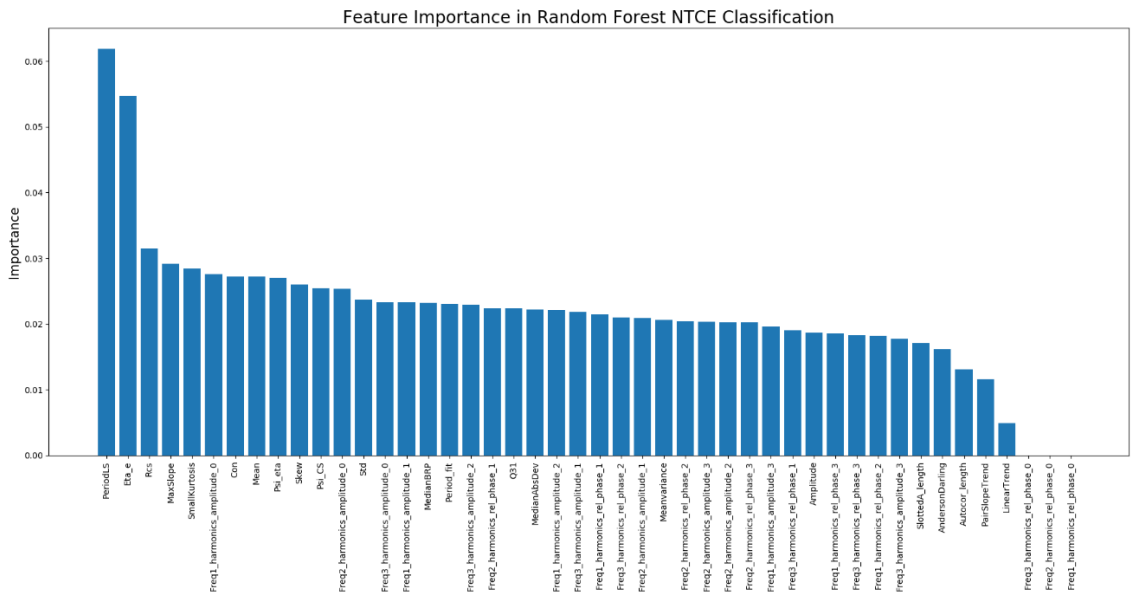

Figure 6 - Feature importance in random forest number of threshold crossing events classification

Using these results we can reduce the data to only the features that hold importance for the classification tasks at hand and improve the overall classification accuracy.

### 3.3.3.2 Classification Results

Upon running each of the five classifier models we are able to see how well each type predicts the correct value based on the two metrics: out of sample error ($E_{out}$), which measures the difference between the expected and empirical error, and Balanced Accuracy. To calculate $E_{out}$ we first used the training data on the scikit learn model. After it was trained, we used it to predict the classes of the testing data. Since the classes of the testing data are known we can compare them to the class that the model predicts. We predicted all of the classes for each data point in the testing data and for every data point misclassified we incremented a counter. After each point's class was predicted, we divided the total number of points in the testing data to get the ratio of points that were misclassified. The calculation for Balanced Accuracy can be seen in section 3.2.3.

Our experiment was to check if the machine learning method could pull from a complete set of noisy/ sparse Kepler data the correct values for each of the classifications described below. For the number of Kepler objects of interest, we had the model classify whether a given star had at least one KOI associated with it. The performance of each of the models for KOI is in Table 7.

| Number of Kepler Objects of Interest | | |
|---|---|---|
| Classifier | $E_{out}$ (%) | Balanced Accuracy (%) |
| Naïve Bayes | 92.38 | 51.15 |
| SVM | 5.71 | 57.72 |
| KNN | 4.72 | 58.96 |
| D-Tree | 8.19 | 62.82 |
| Random Forest (1000 Trees) | 4.64 | 57.58 |

Table 7 – Comparison of classifier models for classifying the number of Kepler objects of interest for a given set of light curves

Here we see that a Naïve Bayesian model has a very high $E_{out}$ and guessing-level balanced accuracy. Each of the others have low out of sample error and the model appears to be moving beyond simply guessing. These values of balanced accuracy are promising.

Turning to the number of threshold crossing events in Table 8 we start to see more favorable values for balanced accuracy. Here the models were ran against the full light curve set to determine if a given star had a transit event.

| Number of Threshold Crossing Events | | |
|---|---|---|
| Classifier | $E_{out}$ (%) | Balanced Accuracy (%) |
| Naïve Bayes | 87.25 | 50.04 |
| SVM | 14.98 | 66.86 |
| KNN | 10.84 | 62.77 |
| D-Tree | 13.99 | 71.53 |
| Random Forest (1000 Trees) | 8.029 | 69.94 |

Table 8 - Comparison of classifier models for classifying the number of threshold crossing events for a given set of light curves

Here, again, we see that a Naïve Bayesian model is not a good classifier for determining the number of threshold crossing events. It has a large out-of-sample error and its balanced accuracy is that of guessing. Each of the other model types are better to varying degrees. Decision trees produced the best balanced accuracy but had a slightly higher out-of-sample error than that of a random forest of 1000 trees which produced the least out-of-sample error and a very respectable value for balanced accuracy. When compared to other examples using noisy sparse data with multi-layer neural networks in the extended physics community the model is classifying fairly well. As an example in a related field where the signal of interest (Higgs Boson) is an extreme minority in the an otherwise large data set, the use of deep neural networks (DNNs) for classifying the Higgs Boson at the LHC (Alves, 2017) achieved accuracies ranging from (~60% to 84%).

Finally we look at the number of confirmed planets per stellar light curve in Table 9. Here we wanted to see if we could move beyond a potential transit and determine if the method could identify planets with some degree of confidence. This would be very attractive as it would allow a set of data to be evaluated against training data containing confirmed planets and confidently tell us that a star that has yet to be analyzed indeed has a planet.

| Number of Confirmed Planets | | |
|---|---|---|
| Classifier | $E_{out}$ (%) | Balanced Accuracy (%) |
| Naïve Bayes | 94.54 | 52.23 |
| SVM | 0.745 | 55 |
| KNN | 0.745 | 55 |
| D-Tree | 1.49 | 54.62 |
| Random Forest (1000 Trees) | .745 | 55 |

Table 9 - Comparison of classifier models for classifying the number of confirmed planets for a given set of light curves

This classification proved to be difficult for all five classifiers. Here SVM, KNN, and Random Forests were all guessing a classification of zero (i.e. no confirmed planets). It only makes sense that the model might revert to this since the sample size is very small for light curves containing confirmed planets and even smaller for light curves with multiple planets.

### 3.3.4 Further Analysis

After achieving the first set of results, we decided to take a look at what might be causing such low model performance. We determined that our classification problems are heavily imbalanced, where the minority class (such as a light curve with a confirmed planet) is significantly less than the majority class (light curves without confirmed planets). To help remedy this we turned to a method to attempt to oversample the minority class. Specifically we used Synthetic Minority Over-sampling Technique, or SMOTE (Chawla et al 2002).

In many cases with real data the interesting examples within the data can be severely underrepresented making classification difficult. The machine learning community has approached this probably through both resampling the original dataset (either by oversampling the minority class or under-sampling the majority class) (Kubat & Matwin 1997, Japkowicz 2000, Lewis & Cattlet 1994, Ling & Li 1998) or by adding costs to the training examples (Pazzani, et al. 1994, Domingos 1999). SMOTE provides an approach that combines both oversampling the minority (or interesting) class and under-sampling the majority class. Chawla et al. 2002 used several different classifiers (C4.5 decision trees (Quinlan 1992), Naïve Bayes, and Ripper (Cohen 1995)) and showed that this combined method achieves better performance.

This algorithm does the following:

1) Takes the minority class sample, $x_i$, and its k minority class nearest neighbors $y_1....y_k$
2) Introduces n synthetic examples along the line segments joining $x_i$ with its k neighbors.
    a. Take the difference between $y_j$ and $x_i$
    b. Multiply difference by number between zero and one
    c. Add the difference to $x_i$

We chose to apply this to our most promising classification set, the number of threshold crossing events. Table 10 shows our post-SMOTE classification ratios where class = 0 represents light curves without a crossing event and class = 1 where a crossing event is detected.

| Post-SMOTE Classification Ratios | | |
|---|---|---|
| SMOTE Model | Class = 0 | Class =1 |
| No SMOTE | 0.8781 | 0.1219 |
| Regular | 0.5 | 0.5 |
| Baseline 1 | 0.5 | 0.5 |
| Baseline 2 | 0.5 | 0.5 |
| SVM | 0.5001 | 0.4999 |

Table 10 – Model results before and after applying SMOTE to balance the majority/minority class.

As one can see, the data is dominated by class = 0 events, but by applying the various versions of a SMOTE model we achieve closer to a 50/50 ratio of class = 0 and class = 1 events. To provide additional metrics for model performance with the addition of SMOTE, we calculated recall and precision. Recall is defined as,

$$Recall = TP/(TP + FN)$$

which is basically the ratio of positives that are correct out of all actual positives, and precision is defined as,

$$Precision = TP/(TP + FP)$$

which is the ratio of positives that are correct out of all guessed positives.

Table 11 shows the previously reported $E_{out}$ and balanced accuracies as well as recall and precision for each model classifier.

| Number of Threshold Crossing Events (NCTE) | | | | | |
|---|---|---|---|---|---|
| Classifier | $E_{out}$ (%) | Balanced Accuracy (%) | Recall | Precision | $F_1$ Score |
| Naïve Bayes | 87.25 | 50.04 | 0.993 | 0.122 | 0.217 |
| SVM | 14.98 | 66.86 | 0.429 | 0.394 | 0.411 |
| KNN | 10.84 | 62.77 | 0.279 | 0.621 | 0.385 |
| D-Tree | 13.99 | 71.53 | 0.517 | 0.409 | 0.457 |
| Random Forest (1000 Trees) | 8.029 | 69.94 | 0.395 | 0.828 | 0.535 |

Table 11 – Comparison of the classifier models including recall and precision as metrics

After applying SMOTE to our data we saw that the SMOTE SVM achieved the greatest improvement in results, which can be seen in Table 12.

| NTCE with SMOTE SVM | | | | | |
|---|---|---|---|---|---|
| Classifier | $E_{out}$ (%) | Balanced Accuracy (%) | Recall | Precision | $F_1$ Score |
| Naïve Bayes | 85.4 | 50.5 | 0.980 | 0.123 | 0.219 |
| SVM | 13.2 | 65.8 | 0.381 | 0.452 | 0.413 |
| KNN | 21.4 | 62.0 | 0.517 | 0.289 | 0.371 |
| D-Tree | 17.1 | 69.8 | 0.524 | 0.362 | 0.428 |
| Random Forest (1000 Trees) | 8.86 | 74.7 | 0.530 | 0.672 | 0.593 |

Table 12 – Comparison of the classifier models after applying SMOTE to the minority class. Notice the improvement in balanced accuracy for the random forest and the trade-off for achieving better recall at the sacrifice of some precision.

The Naïve Bayesian model still produced guessing results with a high $E_{out}$. However, the use of SMOTE greatly improved the random forest, in particular, achieving almost 75% balanced accuracy. Additionally, in all cases but the D-Tree, the F1 Score went up slightly. As a comparison, Armstrong et al. 2017 achieved 87% accuracy with a self-organizing map (SOM) neural network for finding the true planet detections and discarding the false positives among the KOIs. While the methods used between Armstrong 2017 and our work are very different, the results give us an idea about the type of accuracy obtainable with real Kepler data.

With more data containing positive detections and additional data conditioning, a SMOTE feature engineering method could be useful in getting insight into exoplanet presence in a given stellar system. This could provide astronomers a useful tool for quickly identifying stellar systems with an extremely high likelihood of exoplanet presence allowing for more focused analyses.

## 4 DISCUSSION

While the results from the LSTM RNN was initially disappointing, this led us to investigate feature engineering for both regression and classification problems. Feature engineering provided excellent results for regression and very promising results for classification. Once we achieved some confidence in the approach, we decided to employ the SMOTE technique on our data set to remove the severe imbalance. After employing SMOTE the classification results greatly improved. The classification balance accuracy is in line with other results with real data (using different methods) but still does not achieve the ~90-95% being achieved with simulated data (Charnock & Moss, 2017). Noisiness and sparseness of the data appear to play a large role in the ability to classify real light curve data with machine learning techniques, but improvements in performance can be made utilizing minority class oversampling techniques such as SMOTE.

Initially we thought that the poor results from the LSTM RNN were entirely due to the noisiness and sparseness of the data and that light curves may not be suited to analysis with an RNN. However, with the success of SMOTE, we now think that there may be techniques to boost the minority class and potentially improve the performance of representation learning methods with real astrophysical data. It should be noted that SMOTE cannot be used on time series data as it is dependent upon existing in feature space and is not applied to raw time series data. Future work will be to investigate such methods and test if the LSTM RNN can be more successful with data augmentation. This reinvestigation could be complementary to the work by Naul et al. 2018 using RNN feature extraction.

The success of the feature engineering approach (particularly with stellar property prediction) gives us confidence that these techniques will make useful tools for the astronomy community when beginning to analyze the large volume of data that will be available with TESS and JWST and provide better guidance in using precious revisit time from ground-based observatories.

## 5 SUMMARY

With the eminent boom of astronomical data on the horizon, new methods and techniques need to be developed and refined to reduce analysis time, increase accuracy, and provide new insights into the data itself. We attempt to add techniques to the community through investigating representation learning and feature engineering approaches to better understand what may be possible.

Upon investigation, we discovered that our LSTM RNN approach to representation learning was limited in its applicability. This was either due to the limited positive sample size within our data or the sparseness and/or noisiness of real data, since successful applications of RNNs have been shown primarily on simulated data where noise is also simulated and therefore more predictable.

While representation learning did not prove to be ideal, feature engineering provided excellent results with regards to both regression and classification. For regression, the model could predict values for density, stellar radius, and effective temperature where the ridge regression model performed the best with a normalized RMS Error of $\pm\,0.0215$, $\pm\,0.0441$, and $\pm\,0.0347$ for each value respectively. Classification results showed that a random forest of 1000 trees produced the lowest out-of-sample error at 8.86% with a balanced accuracy of 74.7%. Upon inspection of the literature in the community, this may be the first comparative study of machine learning methods using real astronomical data. We hope this work will be informative and provide a base for future endeavors both from our team and the extended community.

## 6 ACKNOWLEDGEMENTS

We would like to thank the Kepler team for making all of the data publicly available on MAST. We would also like to thank Sara Seager (MIT) and David Hogg (NYU) for discussions, encouragement, and advice. This work was supported by Northrop Grumman Corporation.

## REFERENCES

Alves, A., Ghosh, T., Sinha, K. 2017 Phys. Rev D 96, 035022

Armstrong, D.J., Pollacco, D., Santerne, A. 2017 MNRAS, 465, 3

Bailey, S., Aragon, C., Romano, R., Thomas, R.C., Weaver, B.A., Wong, D., 2007, The Astrophysical Journal, 665, 2

Ball, N.M., Brunner, R.J., 2010, Int. J. Mod. Phys. D, 19, 1049

Bastien, F., Stassun, K., Basri, G., Pepper, J. 2015, The Astrophysical Journal, 818, 43

Beichman, Benneke, Knutson, et al. 2014, PASP 126 1134

Bottou, L., Proceedings of COMPSTAT '2010, 2010 pages 177–186

Cabrera-Vives, G., Reyes, I., Forster, F., Estevez, P., Maureira, J.C., 2017, The Astrophysical Journal, 836, 1

Charnock T., Moss, A. 2017, ApJ Letters 837: L28

Chawla, N., Bowyer, K., Hall, L., Kegelmeyer, W., 2002 Journal of Artificial Intelligence 16, 321-257

Cohen, W. W., 1995b Proc. 12th International Conference on Machine Learning, pp. 115–123

Domingos, P., 1999 Proceedings of the Fifth ACM SIGKDD International Conference on Knowledge Discovery and Data Mining, pp. 155–164

Glorot, X., Bordes, A., and Bengio, Y., 2011 Proceedings of the Fourteenth International Conference on Artificial Intelligence and Statistics, pages 315–323, 2011

Japkowicz, N., 2000 Proceedings of the 2000 International Conference on Artificial Intelligence

Special Track on Inductive Learning Las Vegas, Nevada.

Karpenka, N.V., Feroz, F., Hobson, M.P., 2013 MNRAS, 429, 2, p. 1278-1285

Kim, D.-W, Protopapas, P., Alcock, C., Byun, Y.-I., & Bianco, F., 2008 13, 1

Kim, D.-W, Protopapas, P., Bailer-Jones, C. A. L., et al. 2014, 18

Kim, D.-W, Protopapas, C., Byun, Y.-I., et al. 2011, The Astrophysical Journal, 735, 68

Kohavi, R., and Provost, F. 1998. In Editorial for the Special Issue on Applications of Machine Learning and the Knowledge Discovery Process, Columbia University, New York, volume 30.

Kubat, M., & Matwin, S., 1997 Proceedings of the Fourteenth International Conference on Machine Learning, pp. 179–186

Lewis, D., & Catlett, J., 1994 Proceedings of the Eleventh International Conference of Machine Learning, pp.148–156

Ling, C., & Li, C., 1998 Proceedings of the Fourth International Conference on Knowledge Discovery and Data Mining (KDD-98)

Mishkin, D., Sergievskiy, N., and Matas, J. 2016, arXiv preprint arXiv:1606.02228

Murphy, K.P., 2012, 'Machine Learning: A Probabilistic Perspective', The MIT Press

Naul, B., Bloom, J.S., Perez, F., and van der Walt, S., 2018, Nature Astronomy 2, 151-155

Nun I., Protopapas, P., Sim, B., Ming, Z., Rahul, D., Castro, N., Pichara, K., 2015 arXiv: 1510.05988

Pazzani, M., Merz, C., Murphy, P., Ali, K., Hume, T., & Brunk, C., 1994 Proceedings of the Eleventh International Conference on Machine Learning

Quinlan, J., 1992 Programs for Machine Learning

Rasmussen, C.E., Williams, C.K.I., Gaussian Processes for Machine Learning, the MIT Press 2006 ISBN 026218253X

Richards, J.W., Starr, D.L., Butler, N.R., et al. 2011, The Astrophysical Journal, 733, 10

Sak, H., Senior, A., and Beaufays, F. 2014, Fifteenth Annual Conference of the International Speech Communication Association

Schmidhuber, Jurgen, 2015, Neural Networks 61, 85-117

Sharma S., Stello, D., Buder, S., et al. 2017 arXiv: 1709.00794

Spencer, R. IET Seminar on Data Analytics 2013: Deriving Intelligence and Value from Big Data

Thompson, S., Mulally, F., Coughlin, J., et al. 2015, The Astrophysical Journal, 812, 1

Tibshirani, R. 1996, Journal of the Royal Statistical Society. Series B, 58, 1, 267-288

Wang, D., Hogg, D.W., Foreman-Mackey, D., Scholkopf, B. 2016 arXiv: 1508.01853v2

Zhang, W., Itoh, K., Tanida, J., and Ichioka, I., 1990, Applied optics, 29(32):4790–4797